\documentclass[12pt]{article}
\usepackage{amsmath}
\usepackage{graphics}
\usepackage{epstopdf, color}

\begin{document}

\title{Constructions of the soluble potentials for the non-relativistic quantum system by means of the Heun functions}
\author{Shishan Dong$^{1}$, G. Y\'{a}\~{n}ez-Navarro$^{2}$, M. A. Mercado S\'{a}nchez$^{3}$, C. Mej\'{i}a Garc\'{i}a$^{2}$\\
Guo-Hua Sun$^{4}$, and Shi-Hai Dong$^{3}$\thanks{E-mail address: dongsh2@yahoo.  com}\\
{\footnotesize $^{1}$Information and Engineering College, Dalian University, Dalian 116622, P. R. China}\\
{\footnotesize $^{2}$Escuela Superior de F\'isica y Matem\'aticas,  Instituto Polit\'ecnico Nacional,  Edificio 9,}\\
{\footnotesize  Unidad Profesional ALM,  CDMX C.  P. 07738,  Mexico}\\
{\footnotesize $^{3}$Laboratorio de Informaci\'{o}n Cu\'{a}ntica, CIDETEC, Instituto Polit\'ecnico Nacional,}\\
{\footnotesize Unidad Profesional ALM, CDMX 07700, Mexico}\\
{\footnotesize $^{1}$Catedr\'atica CONACyT, CIC, Instituto Polit\'ecnico Nacional, Unidad Profesional ALM, CDMX 07700, Mexico}}

\date{}
\maketitle

\begin{abstract}
The Schr\"{o}dinger equation $\psi''(x)+\kappa^2 \psi(x)=0$ where $\kappa^2=k^2-V(x)$ is rewritten as a more popular form of a second order differential equation through taking a similarity transformation $\psi(z)=\phi(z)u(z)$ with $z=z(x)$. The Schr\"{o}dinger invariant $I_{S}(x)$ can be calculated directly by the Schwarzian derivative $\{z, x\}$ and the invariant $I(z)$ of the differential equation $u_{zz}+f(z)u_{z}+g(z)u=0$. We find an important relation for moving particle as $\nabla^2=-I_{S}(x)$ and thus explain the reason why the Schr\"{o}dinger invariant $I_{S}(x)$ keeps constant. As an illustration, we take the typical Heun differential equation as an object to construct a class of soluble potentials and generalize the previous results through choosing different $\rho=z'(x)$ as before. We get a more general solution $z(x)$ through integrating $(z')^2=\alpha_{1}z^2+\beta_{1}z+\gamma_{1}$ directly and it includes all possibilities for those parameters. Some particular cases are discussed in detail.

\end{abstract}

\vskip 1mm \noindent {\bf Key words}: Schr\"{o}dinger invariant, Schwarzian derivative, Heun differential equation, Soluble potentials.  \\
{\bf{PACS numbers}}: 03.  30. Gp, 02.  30. Hq, 03.  65.  Ge

\section{Introduction}
The exact solution of the Schr\"{o}dinger equation with physical potentials has played an important role in quantum mechanics. Generally speaking, for a given external field, one of our main tasks is to show how to solve the differential equation through choosing suitable variables and then find its solutions can be expressed by some special functions. Here we focus on how to construct a class of the solvable potentials within the framework of the non-relativistic Schr\"{o}dinger equation. Similar works have been carried out \cite{C2, C3, C1, C4, C5}, but it is worth pointing out that Ishkhanyan and his co-authors took $\rho^2\propto (z-a_1)^{m_1}(z-a_2)^{m_2}(z-a_3)^{m_3}$, where the parameters $a_{1, 2, 3}$ are three singularity points, to construct the soluble potentials with some constraints on the parameters $-1\leq m_{1, 2, 3}\leq 1$ and $1\leq m_{1}+m_2+m_3\leq 3$ \cite{C4, C5}. By choosing different values of these parameters which satisfy these constraints, some interesting results have been obtained. However, the approaches which were taken by Natanzon \cite{C3}, who constructed a class of the soluble potentials related to the hypergeometric functions and Bose \cite{C2}, who discussed the Riemann and Whittaker differential equations are different from Ishkhanyan {\it et al}.  Nevertheless, in Bose's classical work \cite{C2} he only studied a few special cases for the differential equation $(z')^2=\alpha_{1}z^2+\beta_{1}z+\gamma_{1}$. Its general solutions were not presented at that time due to the limit on the possible computation condition. In this work our aim is to construct the soluble potentials within the framework of the Schr\"{o}dinger invariant $I_{S}(x)$ through solving $z'(x)$ differential equation directly and then obtaining its more general solutions but not only considering several special cases for the parameters $\alpha_{1}$, $\beta_{1}$ and $\gamma_{1}$.

The rest of this work is organized as follows. In Section 2 we present the Schwarzian derivative $\{z, x\}$ and the invariant $I(z)$ of the differential equation $u_{zz}+f(z)u_{z}+g(z)u=0$ through acting the similarity transformation $\psi(z)=\phi(z)u(z)$ on the Schr\"{o}dinger equation. In Section 3 as an illustration we take the Heun differential equation as a typical example but with different approach taken by Ishkhanyan {\it et al}.  The all soluble potentials are obtained completely in Section 4. Some concluding remarks are given in Section 5.

\section{Similarity transformations to the Schr\"{o}dinger equation}
As we know, the Schr\"{o}dinger equation has the form
\begin{equation}\label{sch}
\frac{d^{2}}{dx^{2}}\psi(x)+[k^2-V(x)]\psi(x)=0.
\end{equation} where we call $k^2$ an energy term and $V(x)$ an external potential.

Through choosing a similarity transformation $\psi(z)=\phi(z)u(z)$ where $z=z(x)$, we are able to obtain the following differential equation
\begin{align}\label{ec1}
u_{zz}(z)+\biggl(\frac{\rho_{z}}{\rho}+\frac{2\phi_{z}}{\phi}\biggr)u_{z}(z)+\biggl[\frac{\phi_{zz}}{\phi}+\frac{\rho_{z}\phi_{z}}{\rho\phi}+\frac{k^2-V(x)}{\rho^{2}}\biggr]u(z)=0, ~~\rho(x)=\frac{dz(x)}{dx},
\end{align}which can be rewritten as
\begin{align}\label{ec2}
u_{zz}+f(z)u_{z}+g(z)u=0,
\end{align}which implies that
\begin{equation}\label{fg}
f(z)=\left(\frac{\rho_{z}}{\rho}+\frac{2\phi_{z}}{\phi}\right), ~~~~g(z)=\frac{\phi_{zz}}{\phi}+\frac{\rho_{z}\phi_{z}}{\rho\phi}+\frac{k^2-V(x)}{\rho^{2}}.
\end{equation}Integrating the first differential equation allows us to obtain
\begin{align}\label{ec3}
\phi(z)& =\rho^{-\frac{1}{2}}e^{\frac{1}{2}\int{f(z)dz}}.
\end{align}
Substitution of this into the second differential equation of Eqs. (\ref{fg}) yields
\begin{align}\label{ec4}
g-\frac{1}{2}f_{z}-\frac{1}{4}f^{2}& =-\frac{1}{2}\Bigl(\frac{\rho_{z}}{\rho}\Bigr)_{z}-\frac{1}{4}\Bigl(\frac{\rho_{z}}{\rho}\Bigr)^{2}+\frac{(E-V)}{\rho^{2}},
\end{align}from which we define the expression \cite{C55}
\begin{align}\label{ec5}
I(z)=g-\frac{f_{z}}{2}-\frac{f^{2}}{4}
\end{align}
as the invariant\footnote{Such a process is known as the normal form of the equation. Equations which have the same normal form are equivalent.  } of Eq.  \eqref{ec2}. Using Schwarzian derivative
\begin{equation}\label{schwa}
\begin{array}{l}
\{z, x\}=\displaystyle\frac{d^2\log z'(x)}{dx^2}-\frac{1}{2}\left(\frac{d\log z'(x)}{dx}\right)^2=\left(\frac{z^{''}(x)}{z^{'}(x)}\right)^{'}-\frac{1}{2}\left[\frac{z^{''}(x)}{z^{'}(x)}\right]^{2}\\[3mm]
~~~~~~~=\displaystyle\frac{z'''(x)}{z'(x)}-\frac{3}{2}\left[\frac{z''(x)}{z'(x)}\right]^2=
\rho\rho_{zz}-\frac{1}{2}\rho_{z}^2,
\end{array}
\end{equation} where we have used the relation $z^{''}(x)=\rho\frac{d}{dx}=\rho\rho_{z}$ and considering equation \eqref{ec5}, then equation \eqref{ec4} can be rewritten as
\begin{align}\label{ec7}
\rho^{2}I(z)+\frac{1}{2}\{z, x\}=k^2-V(x)\equiv I_{S}(x),
\end{align}
where $I_{S}(x)$ is defined as the Schr\"{o}dinger invariant \cite{C2, C3}. Thus, the problem of the construction of the soluble potentials for the original Schr\"{o}dinger equation (\ref{sch}) is solvable on the basis of the functions corresponding to a given $I(z)$ (\ref{ec5}) becomes a problem of deciding transformations $z(x)$ such that the relation $\rho^{2}I(z)+\frac{1}{2}\{z, x\}=z'(x)^2\rho^{2}I(z)+\frac{1}{2}\{z, x\}=I_{S}(x)$ holds. The Schr\"{o}dinger invariant $I_{S}(x)$ is thus characterized by two elements, i.  e. $I(z)$ and the Schwarzian derivative $\{z, x\}$, which is directly related to the function $z(x)$.

\section{Application to Heun differential equation}

The Heun differential equation is given by \cite{C6, C7, C8, C9, C99}
\begin{align}\label{ec8}
u_{zz}+\Biggl(\frac{\gamma}{z}+\frac{\delta}{z-1}+\frac{\epsilon}{z-a}\Biggr)u_{z}+\frac{\alpha \beta z-q}{z(z-1)(z-a)}u=0,
\end{align}
where the parameters satisfy the Fuchsian relation
\begin{align}\label{ec80}
\alpha+\beta+1=\gamma+\delta+\epsilon.
\end{align}

For this equation, the Heun Invariant $I_{h}$ can be calculated as
\begin{align}\label{ec9}
\begin{array}{l}
I_{h}(z)=\displaystyle\frac{\alpha \beta  z-q}{z (z-1) (z-a)}-\frac{1}{4} \left(\frac{\gamma }{z}+\frac{\delta }{z-1}+\frac{\epsilon }{z-a}\right)^2+\displaystyle\frac{1}{2} \left(\frac{\gamma }{z^2}+\frac{\delta }{(1-z)^2}+\frac{\epsilon }{(a-z)^2}\right)\\[3mm]
~~~~~~=\displaystyle\frac{A\, z^{4}+B\, z^{3}+C\, z^{2}+D\, z+F}{z^{2}(z-1)^{2}(z-a)^{2}}.
\end{array}
\end{align}
where the parameters $A, B, C, D, F$ depend on the parameters $a$ and $\alpha, \beta, \gamma, \delta, \epsilon$, i.  e.
\begin{equation}\label{}
\begin{array}{l}
A=\frac{1}{4}[4 \alpha  \beta -(\gamma +\delta +\epsilon -2) (\gamma +\delta +\epsilon )], \\[3mm]
B=\frac{1}{2} \{-2 (a+1) \alpha  \beta +(\gamma +\delta +\epsilon -2) (a \gamma +a \delta +\gamma +\epsilon )-2 q\}\\[3mm]
C=\frac{1}{4} \Big\{a^2 [-(\gamma +\delta -2)] (\gamma +\delta )+a [4 \alpha  \beta -2 \epsilon  (2 \gamma +\delta )-4 \gamma  (\gamma +\delta -2)]\\[3mm]
~~~~~~~~+4 (a+1) q-(\gamma +\epsilon -2) (\gamma +\epsilon )\Big\}, \\[3mm]
D=\frac{1}{2} a [\gamma  (a (\gamma +\delta -2)+\gamma +\epsilon -2)-2 q], \\[3mm]
F=-\frac{1}{4}a^2  \gamma(\gamma -2).
\end{array}
\end{equation}

\section{Soluble potentials constructed by Heun invariant transferred to Schr\"{o}dinger invariant}

Now, let us determine functions $z'(x)$ that can be used to transform $I_{h}$ to $I_{S}$ in order to calculate the Schwarzian derivative $\{z, x\}$. The form of the present invariant $I_{h}(z)$ given in Eq. (\ref{ec9}) suggests us to take the class of functions defined by\footnote{It should be pointed out that present choice is different from previous one \cite{C5}, in which the $\rho^2\propto (z-a_1)^{m_1}(z-a_2)^{m_2}(z-a_3)^{m_3}$ is chosen in order to adapt the mathematical character of the Heun invariant (\ref{ec9}).  }
\begin{equation}\label{summ}
\rho^2=z'(x)^2=\alpha_{1} z(x)^2+\beta_{1} z(x)+\gamma_{1},
\end{equation}where $\alpha_{1}, \beta_{1}, \gamma_{1}$ are arbitrary constants. Such a choice is to make the $\rho^2 I_{h}(z)$ (\ref{ec7}) generate a constant to cancel the energy level term $k^2$. Otherwise, the expansion terms for $\rho^2 I_{h}(z)$ without including a constant will make the $k^2=0$, which means the particle moving in a free field. In terms of this equation (\ref{summ}), one has
\begin{equation}\label{schwarz}
\left\{z, x\right\}=-\frac{\alpha_{1}}{2}-\frac{3}{8}\frac{\beta_{1}^2-4\alpha_{1}\gamma_{1}}{(\alpha_{1} z^2+\beta_{1} z+\gamma_{1})},
\end{equation}where we have used the relation $\left\{z, x\right\}=\rho\rho_{zz}-\frac{1}{2}\rho_{z}^2$ and $\rho=z'(x)$.

To solve equation (\ref{summ}), we first obtain its general solutions for arbitrary parameters. In this case, one has
\begin{equation}\label{solu}
\begin{array}{l}
z(x)^{\pm}=\displaystyle\frac{e^{\sqrt{\alpha _1} \left[-\left(c_1\pm x\right)\right]} \left[\beta _1^2-4 \alpha _1 \gamma _1-2 \beta _1 e^{\sqrt{\alpha _1} \left(c_1\pm x\right)}+e^{2 \sqrt{\alpha _1} \left(c_1\pm x\right)}\right]}{4 \alpha _1}\\[3mm]
~~~~~=-\displaystyle\frac{\beta_{1}}{2\alpha_{1}}+\frac{e^{\mp \sqrt{\alpha_{1}}x}(\beta_{1}^2-4\alpha_{1}\gamma_{1})+e^{\pm \sqrt{\alpha_{1}}x}}{4\alpha_{1}}, \\[3mm]
\end{array}
\end{equation}where $c_{1}$ is an integral constant and we take $c_{1}=0$ for simplicity.

Let us study it in various options for those parameters. If choosing the constants $\alpha_{1}, \beta_{1}, \gamma_{1}$ and the constants of integration suitably, in terms of above results (\ref{solu}) we can obtain their solutions but ignore unimportant integral constants as follows:

\noindent
1)When $\beta_{1}=\gamma_{1}=0$, $\alpha_{1}=4a^2$, we have
\begin{equation}\label{sol-1}
z_{1}^{\pm}(x)=g \exp (\pm 2a\, x), ~~~g\in \rm constant.
\end{equation}

\noindent
2)When $\alpha_{1}=-\beta_{1}$, $\gamma_{1}=0$, $\alpha_{1}=4a^2$, we have
\begin{equation}\label{sol-2}
z_{2}^{a}=\cosh ^2(a\, x), ~~z_{2}^{b}=-\sinh ^2(a\, x).
\end{equation}

\noindent
3)When $\alpha_{1}=-\beta_{1}$, $\gamma_{1}=0$, $\alpha_{1}=-4b^2$, we have
\begin{equation}\label{sol-3}
z_{3}^{a}=\cos^2(b\, x), ~~~~z_{3}^{b}=\sin ^2(b\, x).
\end{equation}

\noindent
4)When $\alpha_{1}=\gamma_{1}=0$, $\beta_{1}=4c$, we have
\begin{equation}\label{sol-4}
z_{4}=c\, x^2, ~~~~c\in \rm constant.
\end{equation}

\noindent
5)When $\alpha_{1}=\beta_{1}=0$, $\gamma_{1}=\sigma^2$, we have
\begin{equation}\label{sol-5}
z_{5}^{\pm}=\pm \sigma\, x.
\end{equation}
It is not difficult to find that the cases 2) and 3) can be obtained each other by considering the relations $\sin(i\, x)=i\sinh(x)$ and $\cos(i\, x)=\cosh(x)$ when $2a$ is replaced by $2ib$.

On the other hand, it was recalled that \cite{C2, C1}
\begin{equation}\label{id-trans}
\left\{z_{t}, x\right\}=\{z, x\}, ~~~~z_{t}\equiv\frac{A_1\, z+B_1}{C_1\, z+D_1},
\end{equation}where $A_1, B_1, C_1, D_1$ are constants but $A_1D_1-B_1C_\not=0$. From Eq.  (\ref{id-trans}), we have
\begin{equation}\label{}
z=\frac{D_1\, z_t-B_1}{A_1-C_1\, z_t}.
\end{equation}

If differentiating $z_{t}$ given in Eq. (\ref{id-trans}) with respect to $x$ and eliminating the variable $z$, one has
\begin{equation}\label{zt-gen}
\begin{array}{l}
z_{t}^{\prime}(x)=\displaystyle\frac{dz_{t}}{dx}=-\displaystyle\frac{(C_1\, z_{t}-A_1)^2}{B_1C_1-A_1D_1}z'(x), \\[3mm]
\left(\displaystyle\frac{dz_{t}}{dx} \right)^2=\displaystyle\frac{(C_1\, z_t-A_1)^2}{(B_1 C_1-A_1 D_1)^2}[\alpha_{1}(B_1-D_1\, z_{t})^2+\beta_{1}(B_1-D_1\, z_{t})(C_1\, z_{t}-A_1)+\gamma_{1}(C_1\, z_{t}-A_1)^2]\\[3mm]
=\displaystyle\frac{(A_1-C_1\, z_{t})^2 (D_1\, z_{t}-B) [\beta_{1}  (A_1-C_1\, z_{t})-\alpha_{1}B_1+\alpha_{1}  D_1\, z_{t}]+\gamma_{1}  (A_1-C_1\, z_{t})^4}{(B_1 C_1-A_1 D_1)^2},
\end{array}
\end{equation}where $z'(x)$ is given by Eq. (\ref{summ}). It is not difficult to see that the solutions of Eq.  (\ref{zt-gen}) are also possible transformations since it is a generalization of Eq.  (\ref{summ}). Up to now, we have found a class of functions for transforming  $I_{h}$ to $I_{S}$. It is worth noting that this class of functions can be characterized differently. We are going to give a useful remark on the $z_{t}'(x)$ given in Eq.  (\ref{zt-gen}). If we use this to calculate the Schr\"{o}dinger invariant $I_{S}(x)$ (\ref{ec7}), then we will find that the soluble potentials would become rather complicated and do not consider this for simplicity, but it should be recognized that the variable $z_{t}$ is just $z(x)$ as given in Eq.  (\ref{summ}).

We are now in the position to construct the simple Schr\"{o}dinger invariants corresponding to the general Heun differential equation Invariant with the aid of the transformation (\ref{schwarz}) we obtained above. First, let us consider the simpler transform (\ref{summ}). Substituting equations (\ref{ec9}) and (\ref{schwarz}) into Eq. (\ref{ec7}) allows us to obtain the following useful Schr\"{o}dinger invariant
\begin{equation}\label{}
\begin{array}{l}
I_{s}=\rho^2 I_{h}+\frac{1}{2}\left\{z, x\right\}\\[3mm]
=(\alpha_{1} z^2+\beta_{1} z+\gamma_{1})\Big\{\displaystyle\frac{\alpha \beta  z-q}{z (z-1) (z-a)}-\frac{1}{4} \left(\frac{\gamma }{z}+\frac{\delta }{z-1}+\frac{\epsilon }{z-a}\right)^2\\[3mm]
~~+\displaystyle\frac{1}{2} \left[\frac{\gamma }{z^2}+\frac{\delta }{(z-1)^2}+\frac{\epsilon }{(z-a)^2}\right]\Big\}-\left(\frac{\alpha_{1}}{4}+\frac{3}{16}\frac{\beta_{1}^2-4\alpha_{1}\gamma_{1}}{\alpha_{1} z^2+\beta_{1} z+\gamma_{1}}\right)\\[5mm]
=(\alpha_{1}z^2+\beta_{1}z+\gamma_{1})\left[\displaystyle\frac{A\, z^{4}+B\, z^{3}+C\, z^{2}+D\, z+F}{z^{2}(z-1)^{2}(z-a)^{2}}\right]-\displaystyle\left(\frac{\alpha_{1}}{4}+\frac{3}{16}\frac{\beta_{1}^2-4\alpha_{1}\gamma_{1}}{\alpha_{1} z^2+\beta_{1} z+\gamma_{1}}\right)
\end{array}
\end{equation}

Let us write down the Schr\"{o}dinger invariants (essentially related to potentials) based on Eqs. (\ref{sol-1}), (\ref{sol-2}), (\ref{sol-3}), (\ref{sol-4}) and (\ref{sol-5}).

\noindent
i)
\begin{equation}\label{}
\begin{array}{l}
I_{S}(H~1^{+})=a^2 \displaystyle\left\{\frac{4 \left[g e^{2 a x} \left(g e^{2 a x} \left(g e^{2 a x} \left(A g e^{2 a x}+B\right)+C\right)+D\right)+F\right]}{\left(a-g e^{2 a x}\right)^2 \left(g e^{2 a x}-1\right)^2}-1\right\}, \\[4mm]
I_{S}(H~1^{-})=a^2 \displaystyle\left\{\frac{4 \left [B g^3 e^{2 a x}+C g^2 e^{4 a x}+D g e^{6 a x}+F e^{8 a x}+A g^4\right]}{\left(e^{2 a x}-g\right)^2 \left(g-a e^{2 a x}\right)^2}-1\right\}.
\end{array}
\end{equation}
\noindent
ii)
\begin{equation}\label{}
\begin{array}{l}
I_{S}(H~2^{a})=-\displaystyle\frac{a^2 \text{csch}^2(2 a x)}{[\cosh^2(a x)-a]^2}
\Big\{3 a^2+(4-16 A) \cosh ^8(a x)-4 (2 a+4 B+1) \cosh ^6(a x)\\[3mm]
~~~~~~~~~~~~~~+[4 a (a+2)-16 C+3] \cosh ^4(a x)-2 [a (2 a+3)+8 D] \cosh ^2(a x)-16 F\Big\}, \\[3mm]
I_{S}(H~2^{b})=-\displaystyle\frac{a^2 \sinh ^4(a x) \tanh ^2(a x)}{4[\sinh^2(a x)+ a]^2} \Big\{\left(3 a^2-16 F\right) \text{csch}^8(a x)+4 (2 a+4 B+1) \text{csch}^2(a x)\\[3mm]
~~~~~~~~~~~~~~+[4 a (a+2)-16 C+3] \text{csch}^4(a x)+2 [a (2 a+3)+8 D] \text{csch}^6(a x)-16 A+4\Big\}.
\end{array}
\end{equation}
\noindent
iii)
\begin{equation}\label{}
\begin{array}{l}
I_{S}(H~3^{a}\left\{3^{b}\right\})=-\displaystyle\frac{b^2 \csc ^2(2 b x)}{[\cos^2(b x)\{\sin^2(b x)\}-a]^2} \Big\{3 a^2+(4-16 A) \cos ^8(b x) \left\{\sin^8 (b x)\right\}\\[3mm]
~~~~~~~~~~~~~~~~~~~-4 (2 a+4 B+1) \cos ^6(b x)\left\{\sin^6 (b x)\right\}+[4 a (a+2)-16 C+3] \cos ^4(b x)\left\{\sin^4 (b x)\right\}\\[3mm]
~~~~~~~~~~~~~~~~~~~-2 [a (2 a+3)+8 D] \cos ^2(b x) \left\{\sin^2 (b x)\right\}-16 F\Big\}.  \\[3mm]
\end{array}
\end{equation}
\noindent
iv)
\begin{equation}\label{}
I_{S}(H~4)=\displaystyle\frac{16 \left\{c x^2 \left[c x^2 \left(c x^2 \left(A c x^2+B\right)+C\right)+D\right]+F\right\}}{4 x^2\left(c x^2-1\right)^2 \left(a-c x^2\right)^2}-\frac{3}{4 x^2}.
\end{equation}
\noindent
v)
\begin{equation}\label{}
\begin{array}{l}
I_{S}(H~5^{\pm})=\displaystyle\frac{\sigma x [\sigma x (\sigma x (A \sigma x\pm B)+C)\pm D]+F}{x^2 (\sigma x\mp 1)^2 (a\mp \sigma x)^2}.
\end{array}
\end{equation}
Here, we have used the symbol $(H~n^{(\pm, a, b)})$ to denote the above invariants, $H$ referring to $I_{h}$ and $n^{(\pm, a, b)}$ to $z_{n}^{(\pm, a, b)}$. Let us analyze these potentials through expanding them as follows:

For the i) case, we have
\begin{equation}\label{}
I_{S}(H~1^{+})=a^2 (4 A-1)+\frac{A_{1}^{+}}{\left(g e^{2 a x}-1\right)^2}+\frac{B_{1}^{+}}{\left(g e^{2 a x}-1\right)}+\frac{C_{1}^{+}}{\left(g e^{2 a x}-a\right)}+\frac{D_{1}^{+}}{\left(g e^{2 a x}-a\right)^2}
\end{equation}
where
\begin{equation}\label{}
\begin{array}{l}
A_{1}^{+}=\displaystyle\frac{4 \left(a^2 A+a^2 B+a^2 C+a^2 D+a^2 F\right)}{(a-1)^2}, \\[3mm]
B_{1}^{+}=\displaystyle\frac{4 \left(4 a^3 A+3 a^3 B+2 a^3 C+a^3 D-2 a^2 A-a^2 B+a^2 D+2 a^2 F\right)}{(a-1)^3}, \\[3mm]
C_{1}^{+}=\displaystyle\frac{4 \left(2 a^6 A-4 a^5 A+a^5 B-3 a^4 B-2 a^3 C-a^3 D-a^2 D-2 a^2 F\right)}{(a-1)^3}, \\[3mm]
D_{1}^{+}=\displaystyle\frac{4 \left(a^6 A+a^5 B+a^4 C+a^3 D+a^2 F\right)}{(a-1)^2}.
\end{array}
\end{equation}

For the ii) case, we have
\begin{equation}\label{}
\begin{array}{l}
I_{S}(H~2^{a})=\displaystyle\frac{A_{2}^{a}\text{csch}^2(2 a x)}{[\cosh^2(a x)-a]^2}+\frac{B_{2}^{a}\text{csch}^2(a x)}{[\cosh^2(a x)-a]^2}-\displaystyle\frac{C_{2}^{a}\coth ^2(a x)}{[\cosh^2(a x)-a]^2}\\[3mm]
~~~~~~~~~~~~~~~+\displaystyle\frac{ D_{2}^{a} \cosh ^2(a x) \coth ^2(a x)}{[\cosh^2(a x)-a]^2}+\displaystyle\frac{E_{2}^{a}\cosh ^4(a x) \coth ^2(a x)}{[\cosh^2(a x)-a]^2}
\end{array}
\end{equation}
where
\begin{equation}\label{}
\begin{array}{l}
A_{2}^{a}=a^2 \left(16 F-3 a^2\right), ~~~B_{2}^{a}=\displaystyle\frac{a^2 [a (2 a+3)+8 D]}{2}, \\[3mm]
C_{2}^{a}=\displaystyle\frac{a^2 [4 a (a+2)-16 C+3]}{4}, ~~~D_{2}^{a}=a^2 (2 a+4 B+1), \\[3mm]
E_{2}^{a}=a^2 (4 A-1).
\end{array}
\end{equation}

For the iii) case, one has
\begin{equation}\label{}
\begin{array}{l}
I_{S}(H~3^{a})=\displaystyle\frac{A_{3}^{a} \csc ^2(2 b x)}{[\cos (b x)^2-a]^2}+\frac{B_{3}^{a}\csc ^2(b x)}{[\cos (b x)^2-a]^2}-\displaystyle\frac{C_{3}^{a}\cot ^2(b x)}{[\cos (b x)^2-a]^2}\\[3mm]
~~~~~~~~~~~~~+\displaystyle\frac{D_{3}^{a} \cos ^2(b x) \cot ^2(b x)}{[\cos (b x)^2-a]^2}+\displaystyle\frac{E_{3}^{a}\cos ^4(b x) \cot ^2(b x)}{[\cos (b x)^2-a]^2}
\end{array}
\end{equation}
where
\begin{equation}\label{}
\begin{array}{l}
A_{3}^{a}=b^2 \left(16 F-3 a^2\right), ~~~~B_{3}^{a}=\displaystyle\frac{b^2 [a (2 a+3)+8 D]}{2}, \\[3mm]
C_{3}^{a}=\displaystyle\frac{b^2 [4 a (2 + a) - 16 C+3]}{4}, ~~~D_{3}^{a}=b^2 (2 a+4 B+1), ~~~E_{3}^{a}=b^2(4 A-1).
\end{array}
\end{equation}

For the special case iv), we have
\begin{equation}\label{}
I_{S}(H~4)=\frac{A_{4}}{x^2}+\frac{B_{4}}{\left(c x^2-a\right)^2}+\frac{C_{4}}{\left(c x^2-a\right)}+\frac{D_{4}}{\left(c x^2-1\right)^2}+\frac{E_{4}}{\left(c x^2-1\right)}
\end{equation}
where
\begin{equation}\label{}
\begin{array}{l}
A_{4}=\displaystyle\frac{16 F-3 a^2}{4 a^2}, ~~~B_{4}=\displaystyle\frac{4 c \left(a^4 A+a^3 B+a^2 C+a D+F\right)}{(a-1)^2 a}, \\[3mm]
C_{4}=\displaystyle\frac{4 c \left(a^5 A-3 a^4 A-2 a^3 B-a^3 C-a^2 C-2 a^2 D-3 a F+F\right)}{(a-1)^3 a^2}, \\[3mm]
D_{4}=\displaystyle\frac{4 c (A+B+C+D+F)}{(a-1)^2 }, \\[3mm]
E_{4}=\displaystyle\frac{4 c (3 a A+2 a B+a C-a F-A+C+2 D+3 F)}{(a-1)^3}.
\end{array}
\end{equation}

For the v) case, one has
\begin{equation}\label{}
I_{S}(H~5^{-})=\displaystyle\frac{A_{3}^{-}}{x}+\frac{B_{3}^{-}}{x^2}+\frac{C_{3}^{-}}{(a+\sigma  x)^2}+\displaystyle\frac{D_{3}^{-}}{(a+\sigma  x)}+\displaystyle\frac{E_{3}^{-}}{(\sigma  x+1)^2}+\displaystyle\frac{F_{3}^{-}}{(\sigma  x+1)}
\end{equation}
where
\begin{equation}\label{}
\begin{array}{l}
A_{5}^{-}=\displaystyle\frac{-\sigma(a D+2 a F+2 F ) }{a^3}, ~~~B_{3}^{-}=\displaystyle\frac{F}{a^2}, \\[3mm]
C_{5}^{-}=\displaystyle\frac{\sigma ^2(a^4 A +a^3 B+a^2 C +a D +F )}{(a-1)^2 a^2}, \\[3mm]
D_{5}^{-}=\displaystyle\frac{\sigma ^2 (2 a^4 A +a^4 B +a^3 B +2 a^3 C +3 a^2 D -a D +4 a F -2 F )}{(a-1)^3 a^3)}, \\[3mm]
E_{5}^{-}=\displaystyle\frac{\sigma ^2(A +B+C +D +F)}{(a-1)^2}, \\[3mm]
F_{5}^{-}=\displaystyle\frac{\sigma ^2(-2 a A -a B +a D +2 a F -B -2 C -3 D -4 F)}{(a-1)^3}.
\end{array}
\end{equation}
Obviously, the potential given in case i) is more complicated than the usual Eckart potential. The potentials discussed in cases ii) and iii) are more complicated
than the first and second type P\"{o}schl-Teller potentials. The potential studied in case iv) is more like the $x^{-2}+x^{-4}$ while the potential given in v) case essentially is the sum of the Coulomb potential plus a centrifugal term.
The other cases such as $I_{S}(H~1^{-}), I_{S}(H~2^{b}), I_{S}(H~3^{b}), I_{S}(H~5^{+})$ have similar properties to their parters.

Now, let us study the wave function. In terms of Eqs.  (\ref{ec3}) and the function
\begin{equation}\label{}
f(z)=\Biggl(\frac{\gamma}{z}+\frac{\delta}{z-1}+\frac{\epsilon}{z-a}\Biggr).
\end{equation} given in (\ref{ec8}), one has the following form
\begin{equation}\label{wave}
\phi(z)=\frac{1}{\sqrt{\rho}}z^{\frac{\gamma}{2}}(z-1)^{\frac{\delta}{2}}(z-a)^{\frac{\epsilon}{2}},
\end{equation}where $\rho$ given in Eq.  (\ref{summ}) depends on the solutions (\ref{solu}), while those particular cases given in Eqs. (\ref{sol-1}) to (\ref{sol-5}). The partial wave function $u(x)$ involved in the whole wave function $\psi(z)=\phi(z)u(z)$ is given by the Heun functions $H_{l}(a, q, \alpha, \beta, \gamma, \delta, \epsilon;z)$.

\section{Concluding remarks}
The Schr\"{o}dinger equation is rewritten as a more popular form of a second order differential equation through taking a similarity transformation. We find that this classical equation is closely related to the Schwarzian derivative and the invariant identity of the differential equation $u_{zz}+f(z)u_{z}+g(z)u=0$. As a typical differential equation, the corresponding mathematical properties of the Heun differential equation are studied.
Before ending this work, we give a useful remark on the Schr\"{o}dinger invariant $I_{S}(x)$. First, let us consider the Schr\"{o}dinger equation (\ref{sch}) and equation (\ref{ec7}). We find that the Schr\"{o}dinger equation can also be rewritten as $\nabla^2\psi(x)=-I_{S}(x)\psi(x)$. Since $\nabla^2$ represents the kinetic term $T$ of the moving particle, it should keep invariant for the same particle. This is also reflection of the conservation of energy $T+V=E$.

\vskip 5mm
\noindent
{\Large \bf Competing Interests}\\
The authors declare that there is no conflict of interests
regarding the publication of this paper.

\vskip 5mm
\noindent
{\Large \bf Acknowledgments}: This work is supported partially by project 20170938-SIP-IPN,
COFAA-IPN, Mexico.

\newpage


\end{document}